# Improved visible to IR image transformation using synthetic data augmentation with cycle-consistent adversarial networks

Kyongsik Yun*, Kevin Yu, Joseph Osborne,
Sarah Eldin, Luan Nguyen, Alexander Huyen, Thomas Lu
Jet Propulsion Laboratory, California Institute of Technology
4800 Oak Grove Drive, Pasadena, CA 91109

## ABSTRACT

Infrared (IR) images are essential to improve the visibility of dark or camouflaged objects. Object recognition and segmentation based on a neural network using IR images provide more accuracy and insight than color visible images. But the bottleneck is the amount of relevant IR images for training. It is difficult to collect real-world IR images for special purposes, including space exploration, military and fire-fighting applications. To solve this problem, we created color visible and IR images using a Unity-based 3D game editor. These synthetically generated color visible and IR images were used to train cycle consistent adversarial networks (CycleGAN) to convert visible images to IR images. CycleGAN has the advantage that it does not require precisely matching visible and IR pairs for transformation training. In this study, we discovered that additional synthetic data can help improve CycleGAN performance. Neural network training using real data (N = 20) performed more accurate transformations than training using real (N = 10) and synthetic (N = 10) data combinations. The result indicates that the synthetic data cannot exceed the quality of the real data. Neural network training using real (N = 10) and synthetic (N = 100) data combinations showed almost the same performance as training using real data (N = 20). At least 10 times more synthetic data than real data is required to achieve the same performance. In summary, CycleGAN is used with synthetic data to improve the IR image conversion performance of visible images.

**Keywords:** image transformation, data augmentation, deep learning, machine learning, computer vision

## 1. INTRODUCTION

Infrared (IR) images are essential for improving the visibility of objects. For example, using an infrared camera in the dark can improve the sight of a firefighter and reduce the risk of slipping, travel, and falling [1,2]. In our previous study, as part of the Assistant for Understanding Data through Reasoning, Extraction, and sYnthesis (AUDREY) project supported by the Department of Homeland Security (DHS) [3,4], we provided an enhanced vision using deep learning-based image transformation from visible to IR images [5–7]. We found that object recognition and segmentation based on the neural network using IR images were more accurate and insightful than color visible images.

However, the bottleneck is the amount of relevant IR images needed for training. Collecting real IR images for special purposes, including space exploration, military, and fire-fighting applications, is a challenge [5,6]. Simple data augmentation techniques (flipping, translating, and rotating) helped improve the classification accuracy of the convolutional neural network when there was a lack of relevant data in the remote sensing scene classification [8]. Other previous studies showed that data augmentation helped generate the most accurate results in classifying urban areas using polarimetric synthetic aperture radar data [9]. Data augmentation also addresses the overfitting problem in machine learning [10].

Previous studies showed that simple data augmentation can be beneficial by modifying the real data (such as flipping, translating, and rotating the image) while maintaining object categories for object recognition [11]. However, the real data itself is fundamentally limited. If the real data does not vary sufficiently, it means that data augmentation may not be helpful. Then we need to consider sophisticated data augmentation such as synthetic data generation. The recent maturation of 3D game engines and computer physics engines enables realistic synthetic images to be generated. Here we simulated the background and target using a 3D game engine. In this paper, we compared the performance of simple data augmentation and synthetic data generation in the problem of visible to IR image transformation.

*kyun@jpl.nasa.gov; phone 1 818 354-1468; fax 1 818 393-6752; jpl.nasa.gov

## 2. SYNTHETIC DATA GENERATION

We created color visible and IR images using the 3D game editor ARMA3 (Bohemia Interactive, Amsterdam, The Netherlands). ARMA3 is an open, realism-based, military tactical video game that delivers realistic synthetic image generation. ARMA3 is based on Real Virtuality 4 3D engine. ARMA3 uses NVIDIA PhysX, an open source real-time physics engine middleware SDK, to enable real-world simulations of vehicles, land, sea and air. We developed a script to work with ARMA3, which spawns random people and vehicles at random locations in a specified location and imports images from multiple angles in a given 3D environment. We can also specify the motion and behavior of the object. For example, a person can walk down a street or direct a truck to run on a specified route. Through this process, we created hundreds of high-quality images in a matter of hours. In reality, it can take days or weeks to take images in the real world. The relative cost of this synthetic data generation process is negligible compared to the potential cost of collecting similar data in real life.

We further processed the synthetic image using a selective Gaussian blur filter with a blur radius of 5 pixels and a maximum delta of 50. The maximum delta is the maximum difference between the pixel value and the surrounding pixel value (value between 0-255). Those pixels with values higher than this max delta value are not blurred (Figure 1).

For pairs of real-visible images and IR images, we used an open-source visible and infrared camouflage video database provided by the Signal Multimedia and Telecommunications Laboratory at the Federal University of Rio de Janeiro [12](http://www02.smt.ufrj.br/~fusion/).

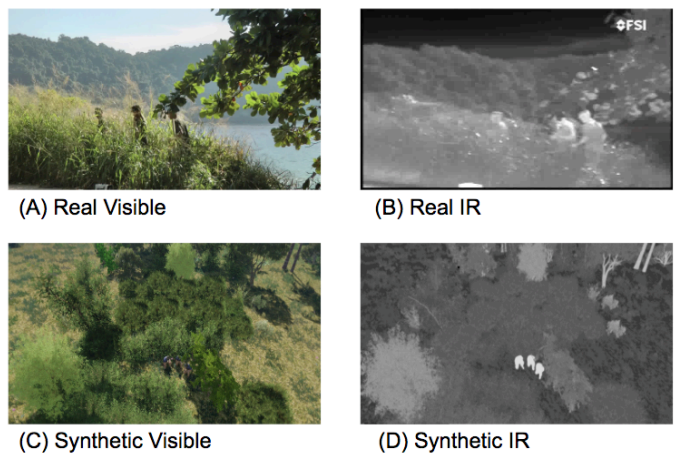

Figure 1. Representative camouflage dataset images of (A) real visible-light, (B) real IR, (C) synthetic visible-light, (D) synthetic IR images.

We also used another dataset that include parking lot surveillance images that contain the visible-light/IR pairs provided by the Army Research Laboratory. The synthetic dataset mimics the situation at various spatial perspectives, different times of the day, and background (Figure 2). In these various situations, creating an actual image can be costly and time-consuming. With a 3D game engine, however, it takes no more than 20 seconds to create a high-quality synthetic image.

With synthetic data, we can pinpoint the exact location of each object (such as a person, car, etc.) based on 3D model information, and place them in any position. We can also shuffle the background and the target to create an infinite combination of synthetic data. For target recognition and segmentation purposes, objects are already fully annotated in the 3D model and do not require manual labeling. In particular, manual labeling for segmentation requires a great deal of time and resources when using real images. Manual labeling by humans may interfere with successful training of neural networks by creating inconsistent ground truth data sets between the labels. The synthetic data solves all of these problems.

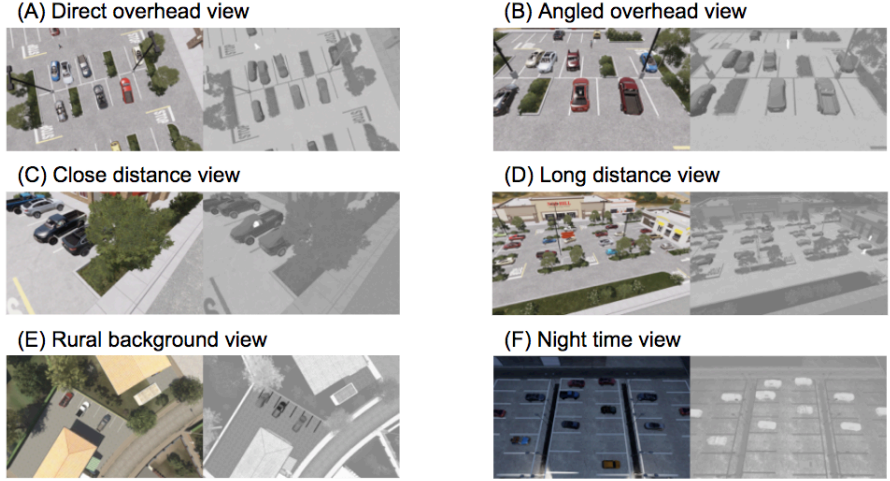

Figure 2. Representative parking-lot synthetic dataset images of (A) direct overhead view, (B) angled overhead view, (C) close distance view, (D) long distance view, (E) rural background view, (F) night time view.

## 3. ADVERSARIAL NETWORKS

We used generative adversarial networks to train associations between various camouflage and parking-lot images of visible-light and IR counterparts. Generative adversarial networks learn the loss of classifying whether the output image is real, while at the same time the networks learn the generative model to minimize this loss [13]. To train the generative adversarial networks, the objective is to solve the min-max game, finding the minimum over $\theta_g$, or the parameters of our generator network G and maximum over $\theta_d$, or the parameters of our discriminator network D.

$$\min_{\theta_g} \max_{\theta_d} \left[ \mathbb{E}_{y \sim p_{data}} \log D_{\theta_d}(y) + \mathbb{E}_{z \sim p(z)} \log \left( 1 - D_{\theta_d}\left( G_{\theta_g}(z) \right) \right) \right] \quad (1)$$

The first term in the objective function (1) is the likelihood, or expectation, of the real data being real from the data distribution $P_{data}$. The log $D(y)$ is the discriminator output for real data $y$. If $y$ is real, $D(y)$ is 1 and log $D(y)$ is 0, which becomes the maximum. The second term is the expectation of $z$ drawn from $P(z)$, meaning the random data input for our generator network (Figure 1). $D(G(z))$ is the output of our discriminator for generated fake data $G(z)$. If $G(z)$ is close to real, $D(G(z))$ is close to 1, and the *log (1-D(G(z)))* becomes very small (minimized).

$$\min_{\theta_g} \max_{\theta_d} \left[ \mathbb{E}_{y \sim p_{data}} \log D_{\theta_d}(y) + \mathbb{E}_{z \sim p(z), x \sim p(x)} \log \left( 1 - D_{\theta_d}\left( G_{\theta_g}(z, x) \right) \right) \right] \quad (2)$$

$$\min_{\theta_g} \max_{\theta_d} \left[ \mathbb{E}_{y \sim p_{data}, x \sim p_{data}} \log D_{\theta_d}(y, x) + \mathbb{E}_{z \sim p(z), x \sim p(x)} \log \left( 1 - D_{\theta_d}\left( G_{\theta_g}(z, x), x \right) \right) \right] \quad (3)$$

We used the random input variable $z$ as input to the generator network. An image can be used as input to the generator network instead of $z$. In the objective function (2), $x$, the real input image was a conditional term for our generator. We then added a conditional term $x$ to our discriminator network, as in function (3). The conditional adversarial networks learned the loss function [14,15] and the mapping from the input image to the output image.

While the original generative adversarial networks generate the real-looking image from the random noise input z, the conditional generative adversarial networks (CGAN) can find the association of the two images (i.e., visible-light and IR). CGAN then converts or reconstructs the real input image into another image. CGAN's one challenge is training data in that they require spatially and temporally matching image pairs for training. The two image spaces have to be closely interrelated. This can be time consuming or even impossible to translate, depending on the two image types. In our case, the visible light and the infrared cameras were not exactly synchronized in time and space. This is where cycle consistent adversarial networks (CycleGAN) is favorable.

With CycleGAN, we can use the deep neural network model to transform between two discrete, unpaired images. They need not be precisely synchronized spatially nor temporally. In order to determine how well the entire translation system performs, CycleGAN introduced two generators (a generator that converts visible-light to IR images, and another generator to convert back from IR to visible-light images). This approach allows CycleGAN to be simultaneously trained in a bidirectional transformation to balance the variability and consistency between training images, resulting in a better translation between unpaired images [16].

Our goal is to translate the captured image in the visible spectrum to the same image captured by the IR wavelength, while at the same time maintaining the structural similarity of the input and output images. We used the U-Net architecture [17] for a progressively downsampled and upsampled encoder-decoder network-based generator for efficiency. We then applied skip layers on the network. That is, each downsampling layer is sent to and coupled to a corresponding upsampling layer. Finally, the upsampling layer can directly learn important structural features in the downsampling layer. For CycleGAN, we used ResNet-based generators to generate high-quality images [16].

In terms of training, all weights were randomly initialized using a Gaussian distribution with mean 0 and standard deviation 0.02. The weight of the discriminator was updated via stochastic gradient descent. Then, the weights of the generators were updated according to the difference between the generated image and the input ground truth. The discriminator and generator ran alternately until convergence or a pre-specified number of epochs passed (epoch = 10,000).

## 4. TRAINING AND TESTING

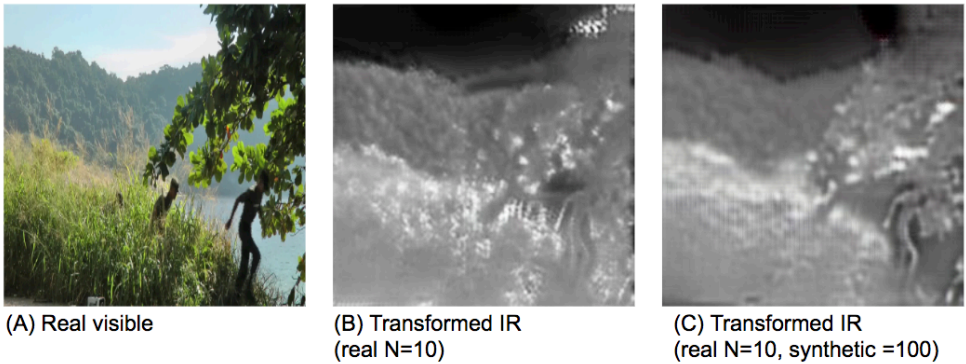

(A) Real visible  (B) Transformed IR (real N=10)  (C) Transformed IR (real N=10, synthetic =100)

Figure 3. Transforming real visible-light images to IR images. (A) Real visible-light image, (B) transformed IR images trained by 10 real images, (C) transformed IR image trained by 10 real and 100 synthetic images. By adding 100 synthetic images for neural network training, the transformation performance improved compared to only 10 real images used for training. Synthetic data helps improve neural network performance.

We found that additional synthetic data helps improve image transformation performance (Figure 3). Neural network training using real data (N = 20) performed more accurate transformations than training using real (N = 10) and synthetic (N = 10) data combinations. Neural network training using real (N = 10) and synthetic (N = 100) data combinations showed almost the same performance as training using real data (N = 20). We used generator L1 loss, which is the difference between generator output and ground truth, as a performance measure (Figure 4).

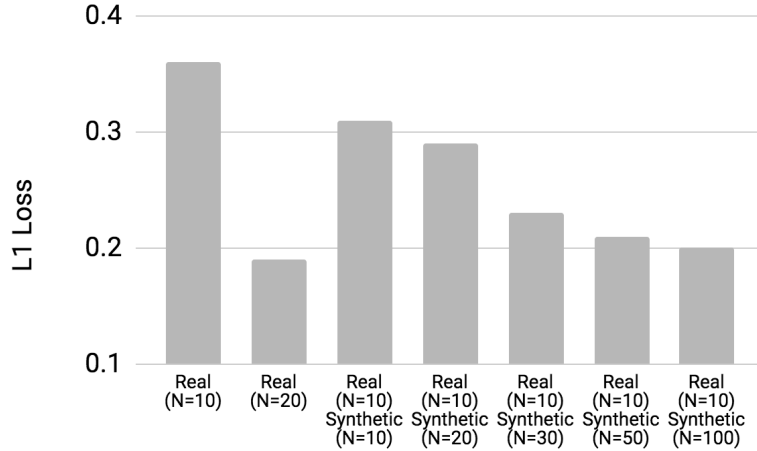

Figure 4. Neural network performance of various data sets. Additional synthetic data helped to improve cycle consistent adversarial network (CycleGAN) performance. Deep neural network training using real data (N = 20) was twice as accurate as training using real (N = 10) and synthetic (N = 10) data combinations. The result indicates that the synthetic data cannot exceed the quality of the real data. The neural network training using real (N = 10) and synthetic (N = 100) data showed almost the same performance as training using real (N = 20) data. This means that more than 10 times as much synthetic data as the real data is required to achieve the same performance.

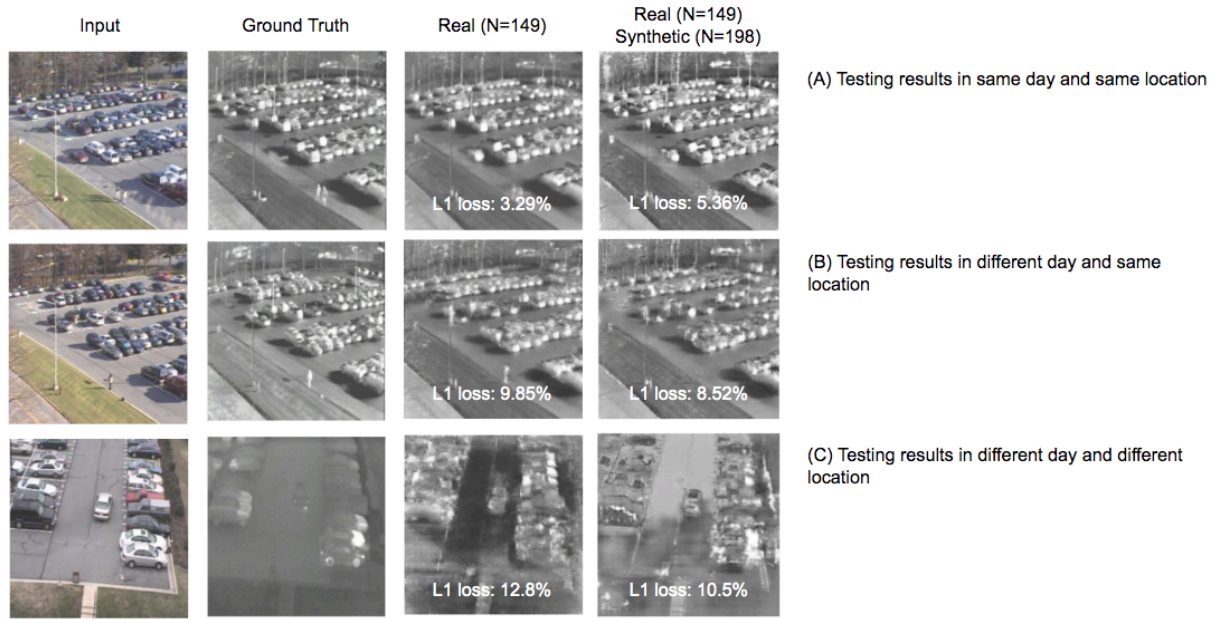

Figure 5. Transformation results of visible-light and IR images. (A) Testing results in same day and same location, (B) testing results in different day and same location, (C) Testing results in different day and different location using real dataset (N=149) and, using combination of real (N=149) and synthetic (N=198) dataset.

A total of 149 real visible-light and IR image pairs in the parking data set were used for training. With limited training data, image transformation was performed relatively well (L1 loss: 3.29%) in a test environment that was consistent with the training environment (same time of a day and same camera perspective). However, the system failed to generalize and yielded poor results in other environments (L1 loss: 12.8%) (Figure 5). Combining the real data with the synthetic dataset (N = 198) improved the test results for the different day (Figure 5B), and for the different day and locations

(Figure 5C), compared to the real image data set. Including synthetic data in the same location on the same day dataset increased L1 loss (Figure 5A). This is because the original real data is overfitted for a particular situation (the same location with the same day). Because synthetic data sets include more diverse environments in terms of time and space in images, training with synthetic data achieved higher performance in a more generalized environment than training with only real data.

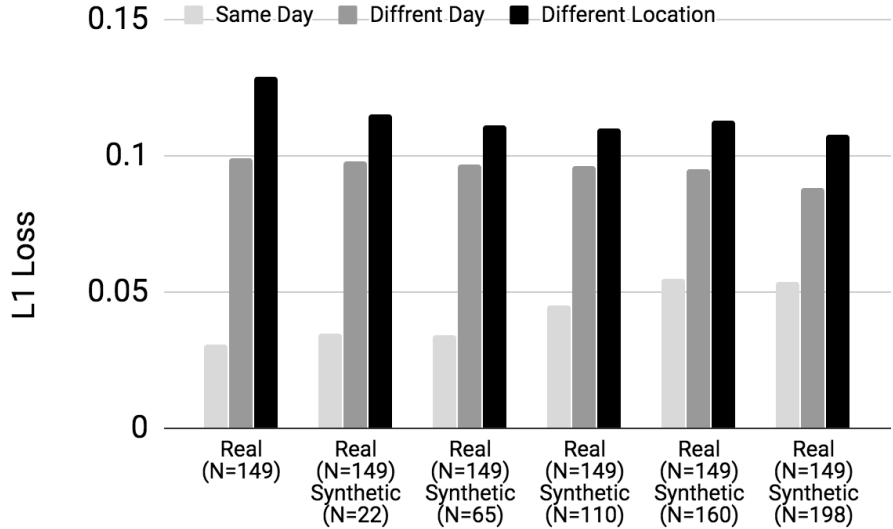

Figure 6. Transformation performance of visible-light and IR images in various conditions. In different day and different location conditions, the L1 loss decreased as we added more synthetic images. In the same day condition, the L1 loss increased as we added more synthetic images that were irrelevant to the purpose of training.

## 5. DISCUSSION

We showed that visible-light images could be transformed to IR images using CycleGAN, and conversion results with limited data sets could be enhanced by adding synthetic data generation based on the 3D game engine. We also found that the amount of synthetic data depends on the quality of the data, but we need at least 10 times more synthetic data than real data to achieve the same performance. Also, if the synthetic data is not relevant to the purpose of the training, the test results may be degraded.

This study used very small data sets, and the generated synthetic data was not very large. Larger synthetic data generation is required for the future research. In a previous 3D pose estimation study, the authors synthesized 2 million images for training to improve performance [18]. Another previous study used data augmentation with manual features (input from experts) to obtain a higher accuracy than the clinical diagnostics in detecting mammographic lesions using a total of 45,000 images [19]. In the medical field, especially large data sets have been used, achieving unprecedented accuracy, which is much higher than the clinical diagnostics [20,21].

We only generated synthetic data using a 3D game engine, and did not perform simple data augmentation techniques, such as flipping, rotating, and zooming in/out. The combination of synthetic data generation and data augmentation may provide more variability in the data set, enabling higher accuracy. Another idea is to change the image style and create a synthetic image using generative adversarial networks. A previous study showed the improvement of classification accuracy by using style transfer with traditional data augmentation [22]. Due to lack of data, generalization was not often done. Data augmentation is known to lead to network regularization and improved test accuracy [23].

In our study, we trained the network from the beginning. The transformation of visible and IR images is a new area that may not be able to benefit from transfer learning (reusing pre-trained models for new problems). However, many previous studies found that deep neural networks could learn transferable features which generalize well to novel tasks [24]. For example, the first three convolutional layers of a pre-trained network can be delivered as a whole to new problems. The next three layers can be transferred and fine-tuned for adaptation. We will benefit from the next version of IR image transformation from visible light through transfer learning, large scale synthetic data generation as well as data augmentation.

## ACKNOWLEDGMENT

The research was carried out at the Jet Propulsion Laboratory, California Institute of Technology, under a contract with the National Aeronautics and Space Administration. The research was funded by the U.S. Department of Homeland Security Science and Technology Directorate Next Generation First Responders Apex Program (DHS S&T NGFR) under NASA prime contract NAS7-03001, Task Plan Number 82-106095.

## REFERENCES

[1] Smith, D. L., [Firefighter fatalities and injuries: the role of heat stress and PPE], Firefighter Life Safety Research Center, Illinois Fire Service Institute, University of Illinois (2008).
[2] Karter, M. J. and Molis, J. L., [US Firefighter Injuries-2012. National Fire Protection Association, Fire Analysis and Research Division, Quincy, MA] (2012).
[3] Sanquist, T. F. and Brisbois, B. R., "Attention and Situational Awareness in First Responder Operations" (2016).
[4] Baggett, R. K. and Foster, C. S., "The Future of Homeland Security Technology," Homeland Security Technologies for the 21st Century **257** (2017).
[5] Yun, K., Bustos, J. and Lu, T., "Predicting Rapid Fire Growth (Flashover) Using Conditional Generative Adversarial Networks," Electronic Imaging **2018**(9), 1–4 (2018).
[6] Yun, K., Lu, T. and Chow, E., "Occluded object reconstruction for first responders with augmented reality glasses using conditional generative adversarial networks," Pattern Recognition and Tracking XXIX **10649**, 106490T, International Society for Optics and Photonics (2018).
[7] Yun, K., Huyen, A. and Lu, T., "Deep Neural Networks for Pattern Recognition," arXiv preprint arXiv:1809.09645 (2018).
[8] Yu, X., Wu, X., Luo, C. and Ren, P., "Deep learning in remote sensing scene classification: a data augmentation enhanced convolutional neural network framework," GIScience & Remote Sensing **54**(5), 741–758 (2017).
[9] De, S., Bruzzone, L., Bhattacharya, A., Bovolo, F. and Chaudhuri, S., "A novel technique based on deep learning and a synthetic target database for classification of urban areas in PolSAR data," IEEE Journal of Selected Topics in Applied Earth Observations and Remote Sensing **11**(1), 154–170 (2018).
[10] Zhang, C., Bengio, S., Hardt, M., Recht, B. and Vinyals, O., "Understanding deep learning requires rethinking generalization," arXiv preprint arXiv:1611.03530 (2016).
[11] Eitel, A., Springenberg, J. T., Spinello, L., Riedmiller, M. and Burgard, W., "Multimodal deep learning for robust RGB-D object recognition," 2015 IEEE/RSJ International Conference on Intelligent Robots and Systems (IROS), 681–687, IEEE (2015).
[12] Ellmauthaler, A., Pagliari, C. L., da Silva, E. A. B., Gois, J. N. and Neves, S. R., "A visible-light and infrared video database for performance evaluation of video/image fusion methods," Multidim Syst Sign Process **30**(1), 119–143 (2019).
[13] Goodfellow, I., Pouget-Abadie, J., Mirza, M., Xu, B., Warde-Farley, D., Ozair, S., Courville, A. and Bengio, Y., "Generative adversarial nets," Advances in neural information processing systems, 2672–2680 (2014).
[14] Isola, P., Zhu, J.-Y., Zhou, T. and Efros, A. A., "Image-to-image translation with conditional adversarial networks," arXiv preprint (2017).
[15] Mirza, M. and Osindero, S., "Conditional generative adversarial nets," arXiv preprint arXiv:1411.1784 (2014).
[16] Zhu, J.-Y., Park, T., Isola, P. and Efros, A. A., "Unpaired image-to-image translation using cycle-consistent adversarial networks," Proceedings of the IEEE International Conference on Computer Vision, 2223–2232 (2017).

[17] Ronneberger, O., Fischer, P. and Brox, T., "U-net: Convolutional networks for biomedical image segmentation," International Conference on Medical image computing and computer-assisted intervention, 234–241, Springer (2015).

[18] Rogez, G. and Schmid, C., "MoCap-guided Data Augmentation for 3D Pose Estimation in the Wild," [Advances in Neural Information Processing Systems 29], D. D. Lee, M. Sugiyama, U. V. Luxburg, I. Guyon, and R. Garnett, Eds., Curran Associates, Inc., 3108–3116 (2016).

[19] Kooi, T., Litjens, G., Van Ginneken, B., Gubern-Mérida, A., Sánchez, C. I., Mann, R., den Heeten, A. and Karssemeijer, N., "Large scale deep learning for computer aided detection of mammographic lesions," Medical image analysis **35**, 303–312 (2017).

[20] Oh, J., Yun, K., Hwang, J.-H. and Chae, J.-H., "classification of suicide attempts through a Machine learning algorithm Based on Multiple systemic Psychiatric scales," Frontiers in psychiatry **8**, 192 (2017).

[21] Litjens, G., Kooi, T., Bejnordi, B. E., Setio, A. A. A., Ciompi, F., Ghafoorian, M., Van Der Laak, J. A., Van Ginneken, B. and Sánchez, C. I., "A survey on deep learning in medical image analysis," Medical image analysis **42**, 60–88 (2017).

[22] Perez, L. and Wang, J., "The Effectiveness of Data Augmentation in Image Classification using Deep Learning," arXiv:1712.04621 [cs] (2017).

[23] Keskar, N. S., Mudigere, D., Nocedal, J., Smelyanskiy, M. and Tang, P. T. P., "On Large-Batch Training for Deep Learning: Generalization Gap and Sharp Minima," arXiv:1609.04836 [cs, math] (2016).

[24] Long, M., Cao, Y., Wang, J. and Jordan, M. I., "Learning transferable features with deep adaptation networks," arXiv preprint arXiv:1502.02791 (2015).